PRE-PRINT ACCEPTED FOR PUBLICATION.
CITATION: *Hill, Benjamin Mako, and Andrés Monroy-Hernández. "The Cost of Collaboration for Code and Art: Evidence from a Remixing Community." In Proceedings of the 2013 ACM Conference on Computer Supported Cooperative Work. San Antonio, TX, USA: ACM 2013.*

# The Cost of Collaboration for Code and Art: Evidence from a Remixing Community

Benjamin Mako Hill (mako@mit.edu)
Andrés Monroy-Hernández (amh@microsoft.com)

**Abstract:** In this paper, we use evidence from a remixing community to evaluate two pieces of common wisdom about collaboration. First, we test the theory that jointly produced works tend to be of higher quality than individually authored products. Second, we test the theory that collaboration improves the quality of functional works like code, but that it works less well for artistic works like images and sounds. We use data from Scratch, a large online community where hundreds of thousands of young users share and remix millions of animations and interactive games. Using peer-ratings as a measure of quality, we estimate a series of fitted regression models and find that collaborative Scratch projects tend to receive ratings that are lower than individually authored works. We also find that code-intensive collaborations are rated higher than media-intensive efforts. We conclude by discussing the limitations and implications of these findings.

## INTRODUCTION

Over the last decade, researchers and theorists studying computer-supported cooperation have pointed to FLOSS, Wikipedia, and the products of remixing communities as examples of how wide scale collaboration facilitated by the Internet – "peer production" – has produced creative works of enormous value and high quality [4, 41]. A subset of this research has focused on remixing: the construction of creative products based on works created previously by others, often involving the recombination and reworking of code with images and sound. Benkler has argued that remixing is an important modality of peer production [4]. In his eponymous book on the subject, Lessig suggests that widespread remixing on the Internet represents an important cultural shift toward a more collaborative culture [29]. Broadly considered, the pattern of indirect reuse central to remixing is characteristic of peer production and a large body of Internet-based collaborative work.

Others have countered that the products of peer production – and remixing communities in particular – are amateurish and of poor quality [27, 22]. Acknowledging the successful examples of GNU/Linux and Wikipedia, Benkler has advanced a nuanced argument that peer production may be particularly well suited to the creation of functional works, like operating system software, "because software is a functional good with measurable qualities" [4]. On the other hand, the last half decade has witnessed high quality peer produced artistic works such as the Grammy-nominated Johnny Cash Project music video.

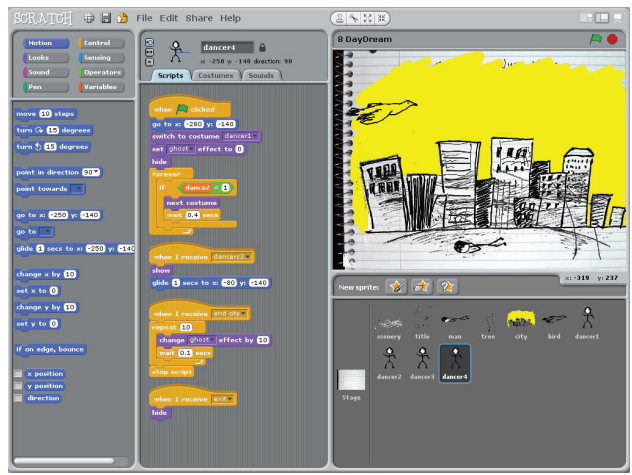

Figure 1: Screenshot of the Scratch desktop application. The leftmost column represents an inventory of "programming blocks" (shown in blue) which have been assembled into program code in the center column. The area in the top right represents the project as it will be displayed to a user interacting with the finished product. The bottom right column shows the available images that can be controlled by the code.

Do remixes result in higher quality creative works, on average, than the products of solitary creators? Is remixing better for functional works like code than it is for creative works like art? Although these questions lie at the core of conversations about peer production, its value, and its limitations, empirical evidence has largely been missing. Internet-based remixing platforms often combine the production of both code and art, and as a result, provide a setting uniquely well suited to answering these questions.

Using peer ratings we seek the answers to these questions in Scratch – a large online remixing community where young people build, share, and collaborate on interactive animations and video games. We find that, on average, remixed projects are rated lower by members of the Scratch community than projects that are the work of a single creator. Additionally, we find that remixes involving artistic creations are less positively evaluated than programmatic works.

## BACKGROUND

The possibility or promise of high quality products, like FLOSS and Wikipedia, has been central to interest in peer production [4, 29, 41]. Benkler has argued that peer production results in high quality projects because the Internet has lowered costs associated with contribution, attracting many lightly motivated contributors [4]. Discussing FLOSS, Raymond suggests that contributors will be able to find and fix shortcomings to produce software that is superior to alternatives produced by individuals or firms [35]. Moreover, there is empirical evidence that supports the argument that increased collaboration leads to higher-quality products in peer production. For example, studies of Wikipedia have suggested that vandalism is detected and removed within minutes or seconds [40] and that high quality articles in Wikipedia, by several measures, tend to be produced by more intense collaboration [43, 23, 12].

That said, we also know that collaborative work is not always better. To cite just one influential example, the rich literature on brainstorming has suggested that, in many cases, groups produce fewer, lower quality, ideas than the same number of individuals working alone [9]. In a baccalaureate speech, Yale President A. Whitney Griswold once expressed this sentiment: "Could Hamlet have been written by a committee, or the Mona Lisa painted by a club? Could the New Testament have been composed as a conference report? Creative ideas do not spring from groups. They spring from individuals. The divine spark leaps from the finger of God to the finger of Adam" [13].

In the literature on peer production and remixing, some have expressed skepticism that increased collaboration will result in better products – the strongest criticisms calling into question the fundamental promise of the phenomena. Keen has argued that the openness of peer production and remixing to participation leads to products that are amateurish and of low quality [22]. In a critique of what he calls "online collectivism," Lanier points to several shortcomings in peer production and suggests that Wikipedia is sterile and lacks a single voice: "[A] voice should be sensed as a whole. You have to have a chance to sense personality for language to have its full meaning. Personal Web pages do that, as do journals and books" [26]. Echoing Griswold, Lanier points out that even in successful collective processes, like scientific endeavors, it is, "individual scholars [that] matter, not just the process or the collective." Previous studies of remixing have shown that factors associated with increased remixing of projects are associated with "cheaper" and easier forms of work, perhaps by lowering costs and attracting dabblers whose contributions involve less complexity and effort [16].

But although remixing and peer production have been the subject of a growing body of research, few studies have attempted to compare the quality of resulting remixes and none has questioned the assumption, fundamental to much of the interest in peer production, that collaboration in the form of remixing leads to higher-quality works. This leads us to our first research question:

> *RQ1: Are remixes, on average, higher quality than single-authored works?*

One response to the critiques of Keen, Lanier, and Griswold is to suggest the peer production *can* result in high-quality artifacts, but only for certain types of creative goods. After all, Griswold's archetypes of works that cannot be created by a committee are all from the arts.

Of course, many great works of art are produced by multiple authors. The King James translation of the Bible was, literally, written by a committee. Even singularly authored works, like the poetry and stories of T.S. Eliot and Raymond Carver, involved deep collaboration with influential 4editors [15]. Research from the psychology literature on creativity has shown that creativity is highly embedded in social contexts [2, 8].

And although peer production's functional products are more widely known, high quality art has also been peer produced. The Johnny Cash Project [21] aggregated thousands of peer produced drawings to create a Grammy-nominated music video. Work by Luther [30] has shown how artistic work is produced in vibrant remixing communities. Recent work looking at the design of virtual objects has shown that remix-style collaboration leads to better designs than when individuals work alone [44]. If artistic products are less visible in peer production and remixing, it might be because the tools necessary to collaborate over the Internet have, historically, been better suited for producing technical products like programming code. New collaborative media and design tools produced over the last decade have shifted that *status quo*.

But, it is undeniable that the poster children of peer production are more "functional" than Griswold's archetypes. For example, Nature's widely cited study evaluated Wikipedia's quality by counting factual errors [11]. Raymond speaks of additional contributors revealing and fixing "bugs" [35]. Works of art may be more or less positively received, but their evaluation is more subjective. Computing researchers have suggested that supporting group work in creative tasks like design or art remains a major challenge and an open area for research [39, 10, 32].

In general, crowdsourcing systems have tended to focus on "less creative" tasks that can be easily divided and objectively evaluated – qualities often missing from creative works like media and film [31]. This can make artistic creations more difficult to manage [30]. Even successful crowd-produced art like Sheep Market [24] and the Johnny Cash Project asked participants to contribute a small well-defined piece of an individual's broader vision. Lanier argues that collectives are more effective than individuals at tasks like setting market prices, but less effective at others, saying, "when a collective designs a product, you get design by committee, a derogatory expression for a reason" [26]. Keen is also doubtful of collective writing and design when he asks, "can a collaboration of amateur voices create an authoritative, coherent fictional narrative?" His answer: "I doubt it" [22].

That said, previous work has struggled to compare artistic and more functional products in a single setting or community. Peer production may be able to create high-quality art, while performing less well than it would with a more functional goal. Additionally, remixers may be able to create great works of art, but show more interest in contributing to more functional tasks when it comes time to collaborate because it leads to smoother social interactions. This conflicting evidence leads us to our second research question:

> *RQ2: Are code-intensive remixes, on average, higher quality than media-intensive remixes?*

Of course, operationalizing and quantifying quality – especially concerning subjective products like art – is difficult and controversial. Some studies on peer production have suggested simple popularity metrics, like views or downloads, as proxies for quality [7]. Others have taken advantage

git revision 4ac6342 on 2012/12/12

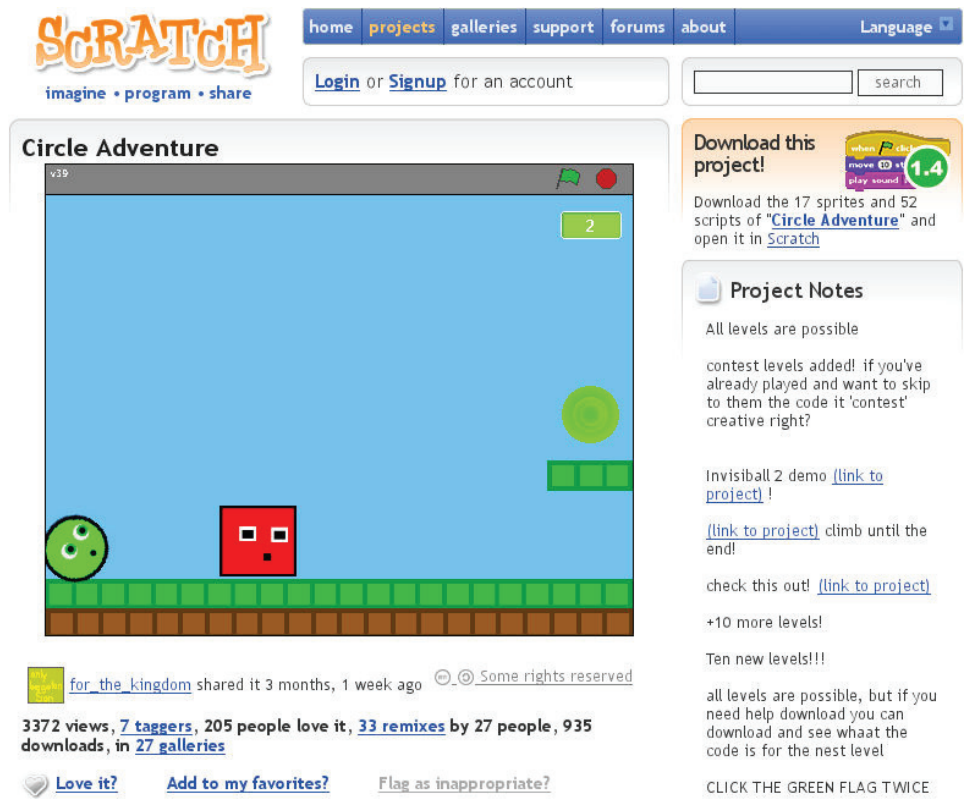

Figure 2: Example of a web page for a single Scratch project. (Accessed: May 29, 2012)

of peer rating systems to help evaluate and rate content [25]. We will follow the latter, rating-based, approach. That said, in results that are not presented here, we experimented with similar analyses using popularity-based quality measures and found substantively similar results.

## SCRATCH

To answer our research questions, we look to the Scratch online community: a public and freely accessible website where a large community of users create, share, and remix interactive media. The community was built to support users of the Scratch programming environment: a freely downloadable desktop application with functionality similar to Adobe Flash [36]. A screenshot of the Scratch programming environment is shown in Figure 1. Crucial to our analysis, Scratch is designed to allow users to build projects integrating images, music, sound and other media with programming code [36]. Scratch was designed by the Lifelong Kindergarten Group at the MIT Media Lab as a platform for creative learning, and aims to introduce young people to computer programming. Much of the practice, and promise, of peer production and remixing has been located in communities of young technology users [29, 19, 33], which makes Scratch an ideal platform to study.

From within the Scratch authoring environment, creators can publish their projects on the

| Variable | N | Mean | SD | Min | Max |
|---|---|---|---|---|---|
| **Dependent Variables** | | | | | |
| Loveits within 1 yr. ($loveits_p$) | 1,271,085 | 0.64 | 6.53 | 0 | 1,320 |
| **Question Predictors** | | | | | |
| Number of blocks ($blocks_p$) | 1,271,085 | 110.52 | 573.45 | 0 | 25,2237 |
| Sum of images and sounds ($media_p$) | 1,271,085 | 22.88 | 70.03 | 1 | 11,624.00 |
| Remix status ($isremix_p$) | 1,271,085 | 0.17 | 0.37 | 0 | 1 |
| **Controls** | | | | | |
| User is female ($female_u$) | 1,270,519 | 0.34 | 0.48 | 0 | 1 |
| User age in years ($age_{up}$) | 1,240,014 | 16.46 | 10.66 | 4 | 74.92 |
| Account age in months ($joined_{up}$) | 1,271,085 | 4.17 | 6.43 | 0 | 49.37 |
| Users' previous loveits ($prevlovits_{up}$) | 1,271,085 | 29.29 | 146.79 | 0 | 1,2287 |
| Views within 1 yr. ($views_p$) | 1,271,085 | 14.17 | 69.26 | 0 | 1,1287 |

Table 1: Summary statistics for variables used in our analysis. Measures with the subscript $p$ are measured at the level of the project while measures with the subscript $u$ are measured at the level of the user. Measures with both are measured at the level of the user at the time that the project was shared.

Scratch online community website hosted at MIT.[1] As of May 2012, more than one million users had created accounts on the website and more than one-third of these users had shared at least one of more than 2.5 million total projects. Any web visitor can create an account, which becomes active immediately. The nature of Scratch projects shared in the community varies widely and includes everything from interactive greeting cards to fractal simulations to animated stories and video games. Projects on the Scratch website vary enormously in complexity in terms of both code and media.

The community is visited by more than half a million people each month[2] who can browse material on the website and interact with projects directly from their web browsers. Visitors must create accounts to remix projects or contribute by publishing, commenting, showing support (i.e., giving "loveits"), tagging, or flagging projects as inappropriate. Most of the community's users self-report their ages ranging between 8 and 17 with 13 being the median age for new accounts. Thirty-five percent of users of the online community self-report as female.

Central to the purposes of our study, the Scratch online community is designed as a platform for collaboration through remixing. Influenced by theories of constructionist learning in communities [34] and communities of practice [28], the Scratch website seeks to help users learn through exposure to, and engagement with, others' projects. The commitment to remixing is deep and visible in Scratch. The name "Scratch" is a reference to hip-hop DJs' practice of remixing. Every project shared on Scratch is available for download and remix by any other user, through a promi-nent download button. Additionally, every project is licensed under a Creative Commons license, explained in "kid friendly terms," that explicitly allows reuse. Administrators and community members actively encourage remixing.[3]

---

[1] http://scratch.mit.edu
[2] http://quantcast.com/scratch.mit.edu

Remixing in Scratch represents a wide range of collaborative modalities, from experimentation, to bug-fixing, to socialization, to coordination within semi-formal organizations, to meme-spreading. For example, organized groups with formal membership use remixing to build projects serially over extended periods of time [3]. Although we lack good knowledge of the distribution of types of remixing, we believe that Scratch remixes include a broad range of the types of cooperative information production studied by human-computer interaction researchers.

DATA AND MEASURES

The Scratch online community is built on top of a database-driven web application that stores an extensive range of metadata on projects, users, and interactions on the website. This database also identifies, tracks, and presents data on whether projects are created through remixing. Additionally, the website stores each of the "raw" Scratch project files, which can be further analyzed to reveal details, such as projects' programming code and media elements. Our dataset is constructed by combining exported metadata about Scratch's users, projects, and interactions, with algorithmic analyses of each project to measure complexity in both code and media.

Our unit of analysis is the Scratch project and our dataset includes every project shared on the Scratch online community from the moment the first project was shared in Scratch on March 5, 2007 through April 1, 2011 – a total of 1,685,353 projects. We omit 19,796 projects for which we do not have data on the amount of code or media because of technical errors in our analytic tools or because of corruption in the project files. We also omit 395,728 projects (24% of the total) which have been removed from the public website by the author. In robustness checks not reported in this paper, we include these hidden projects in our analyses and find that our results are unchanged.

To operationalize quality we use user ratings. In particular, we use a count of *loveits* which are analogous to "liking" or "upvoting" in other social media platforms and which are a indication that a user with an account appreciates a project. Each user can only "love" a project once and can remove her "loveit" from a project later. Users cannot publicly express dislike with a similar mechanism. Loveits are public in the sense that the author of a Scratch project can see which users "loved" her own projects and in the sense that the total number of *loveits* a project has received is visible on the Scratch website next to each project (see Figure 2). This measure of quality is in line with previous work that has suggested the validity of peer rating systems [25] as well as with studies that have used crowd workers for measuring subjective qualities of creative work such as visualizations [14] and photos, [5] which have been shown to be reliable in those contexts.

---

[3] http://info.scratch.mit.edu/License_to_play

We operationalize functional complexity of projects as the number of programming *blocks*. Blocks, also shown in Figure 1, are analogous to tokens or symbols in the source code of computer programs. Blocks are similar to, but more granular than, source lines of code, which have a long history of use as a measure of both complexity and effort in software engineering [42, 1]. We operationalize the artistic intensity of projects as the sum of the number of distinct images and sounds (*media*).

We operationalize collaborativeness using a dummy variable, *isremix*, that indicates whether a project is, itself, a remix of another project. Although this is a crude measure of collaborativeness, only 10.9% of projects are remixed at all and only 1.4% are remixed more than once. As a result, this dichotomous measure captures most variation in collaborativeness in Scratch. It is important to note that this variable only measures material being copied and, as a result, does not capture "conceptual" remixing such as employing a Disney or Nintendo character in a new Scratch game, borrowing material from any source outside Scratch, or inspiration from sources inside Scratch without downloading and copying a project.

We also include a range of controls. For each user, we include self-reported measures of gender, which we have coded as a dichotomous variable (*female*); date of birth, which we have coded as age in years at the moment that each project was shared (*age*), which may indicate sophistication of the user; and the age of each account (*joined*), which may indicate a user's level of experience with Scratch. We are concerned that the prominence of projects' creators may drive positive ratings. To control for prominence, we employ a count the total previous loveits that a project's author had received (*prevloveits*) at the moment that the project in question was shared. The variable is measured at the project level, but is aggregated across all projects previously shared by a user when a user shares the project in question.[4]

Finally, depending on how we conceptualize quality, the popularity of projects may represent an important confounding variable. Our measure of exposure, *views*, reflects the number of visits to a project by logged-in users on the Scratch website. Scratch projects are visible publicly to individuals without accounts but these visits are not recorded and do not count in the tabulation of views. Views, like loveits, are publicly displayed on the Scratch website under each project (see Figure 2). In addition to viewing, logged-in users can download projects to their own computers. In robustness checks not presented here, we found that *downloads* had a similar, but weaker, moderating effect as our control for *views* with an identical pattern of results. Summary statistics for each of our measures are shown in Table 1.

---

[4]Other measures of prominence include number of past views, number of past downloads, and number of past favorites. All are highly correlated with past loveits ($0.89 < \rho < 0.97$) and our results remain unchanged if we use these alternative indicators.

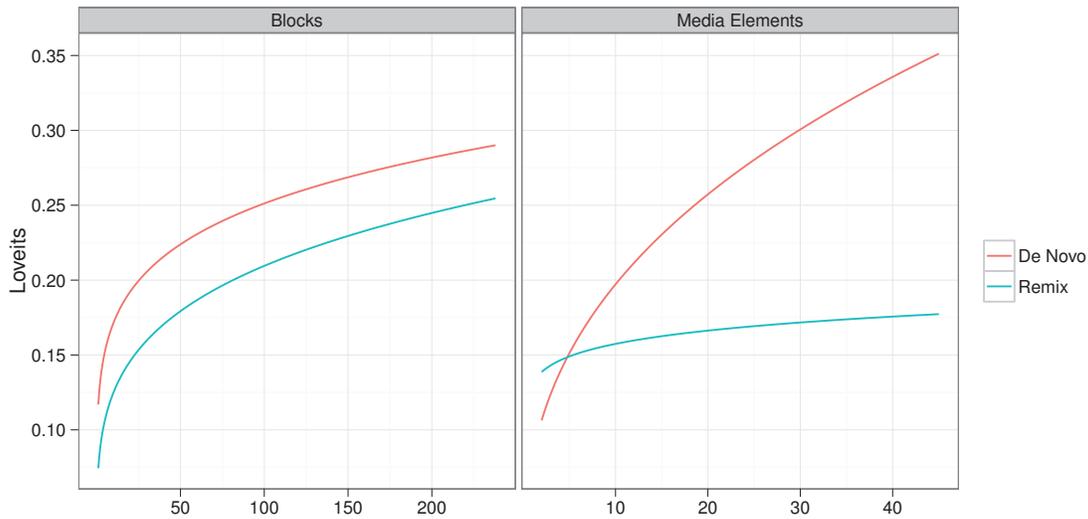

Figure 3: Two plots of estimated values for prototypical projects showing the predicted number of loveits using our estimates in Model 1. In the left panel, the x-axis varies number of blocks while holding media intensity at the sample median. The right panel varies the number of media elements while holding the number of blocks at the sample median. Ranges for each are from 0 to the 90$^{\text{th}}$ percentile.

## ANALYSIS

Our analytic strategy involves the estimation of fitted regression models on our count of *loveits* one year after creation. In both the models we present, we include interaction terms between *isremix* and our measures of project complexity *code* and *media*. Both complexity measures, as well as our measure of prominence (*prevloveits*) are highly skewed, but a started log transformation results in an approximately normal distribution in each case.

One of the challenges of using loveits as a measure of quality is that it may simply measure exposure. Of course, many social media systems use quality measures that are closely related to, or reflect, popularity. That said, more popular projects may receive more loveits only because they are more exposed. As a result, controversial projects, or those reusing elements taken from popular culture, might be viewed more frequently, and receive more loveits, despite the possibility that they are otherwise of lower quality.

We can address this concern by controlling for *views*, our measure of exposure. In Model 1, we do not include a measure of views. These results might be more easily comparable to metrics used in many social media systems. We believe that Model 2, which controls for exposure, offers more conservative estimates. Except for the fact that Model 2 adds the control for *views*, the two models are identical.

Our formal model is shown below and includes the full set of variables including *views*:

$$loveits = \beta_0 + \beta_1 \log blocks + \beta_2 \log media + \beta_3 isremix + \beta_4 \log blocks \times isremix +$$
$$\beta_5 \log media \times isremix + \beta_6 age + \beta_7 joined + \beta_8 female + \beta_9 \log prevloveits + \beta_{10} \log views$$

Poisson regression is frequently used for count dependent variables like *loveits* but, as is common with counts, there is an over-dispersion of zeros in our measure. To address this over-dispersion, we use negative binomial regression to estimate both models.

Because users can upload several projects, and because many do (51% of the users in our dataset), there are concerns about correlation of standard errors between projects uploaded by the same user. We address this by fitting hierarchical mixed-effect models that estimate random intercepts for each user and cluster variance within and between users. These estimates are not reported here but are similar in the magnitude and hypothesis tests.

## RESULTS

Our results are shown in Table 2. The median number of loveits a Scratch project receives after one year is 0 and more than 80% of projects do not receive a single loveit in their first year. As a result, most of the estimated effect sizes in our fitted models represent only fractions of a loveit.

In both models, the principal effects of *blocks* and *media* are positive and statistically significant. In other words, for *de novo* projects, increased complexity in either code or media is, on average, associated with more loveits. This finding is in line with previous work that has suggested that increased complexity of projects is usually associated with positive reception in remixing communities [17, 16]. For example, our Model 1 estimates that a *de novo* project with 3 blocks (10[th] percentile) would be expected to receive, on average, 0.15 loveits; a similar project with 237 blocks (90[th] percentile) would be expected to receive 0.29 loveits. The relationship between *media* and our dependent variables follows a similar pattern.

Our parameter estimate for *isremix* is positive, large, and statistically significant in Model 1, but is negative and statistically significant when we control for *views* in Model 2. That said, this coefficient should not be interpreted directly because the parameter estimates for the interaction terms between *isremix* and both measures of complexity are also statistically significant. To answer our research questions, we must consider the sum of all these effects. Because interaction effects can be difficult to interpret, plots of predicted values for prototypical individuals using the parameter estimates for Model 1, with unvarying qualities held at the sample medians, are shown in Figure 3. Prototypical plots for Model 2 are not included but show similar relationships.

Answering RQ1, our results suggest that most remixes in Scratch will be less highly rated than similar *de novo* projects. Although the principal effects of *isremixed* are positive in both models, the strong negative effect of the interaction between log *media* and *isremix* tends to overwhelm this effect. Except among the least media-intensive projects (see the far left of the right panel in Figure 3), remixes with median values of all other predictors receive less loveits, on average, than otherwise similar *de novo* projects.

|                          | Model 1   | Model 2   |
|--------------------------|-----------|-----------|
| (Intercept)              | −3.359*   | −4.828*   |
|                          | (0.010)   | (0.008)   |
| log *blocks*             | 0.166*    | 0.076*    |
|                          | (0.002)   | (0.002)   |
| log *media*              | 0.384*    | 0.086*    |
|                          | (0.004)   | (0.003)   |
| log *isremix*            | 0.279*    | −0.091*   |
|                          | (0.020)   | (0.014)   |
| *female*                 | 0.000     | 0.284*    |
|                          | (0.005)   | (0.004)   |
| *age*                    | −0.006*   | −0.002*   |
|                          | (0.000)   | (0.000)   |
| *joined*                 | −0.020*   | −0.008*   |
|                          | (0.000)   | (0.000)   |
| *prevloveits*            | 0.568*    | 0.076*    |
|                          | (0.002)   | (0.001)   |
| log *blocks* × *isremix* | 0.059*    | 0.021*    |
|                          | (0.005)   | (0.004)   |
| log *media* × *isremix*  | −0.305*   | −0.103*   |
|                          | (0.009)   | (0.006)   |
| *views*                  |           | 1.226*    |
|                          |           | (0.002)   |
| θ                        | 0.250*    | 1.722*    |
|                          | (0.001)   | (0.011)   |
| N                        | 1239470   | 1239470   |
| AIC                      | 1803040   | 1415524   |
| BIC                      | 1803570   | 1416101   |
| log L                    | -901476   | -707714   |

Standard errors in parentheses;

\** indicates significance at $p < .01$

Table 2: Table of fitted negative binomial regression models estimating a count the number of "loveits" a Scratch project has received one year after being shared on the website.

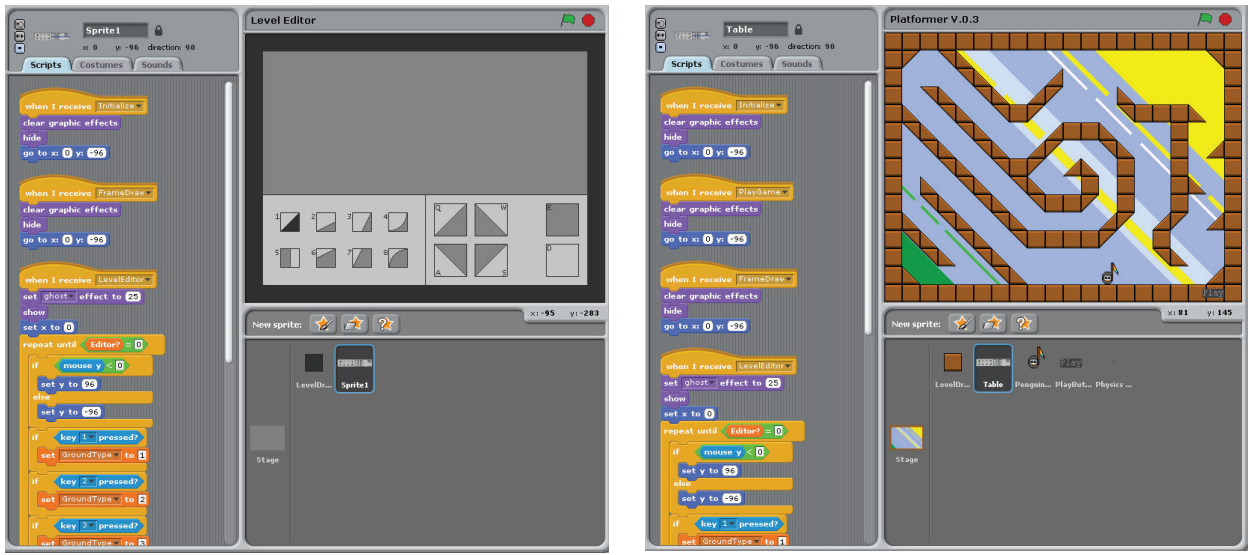

Figure 4: Example of a code-intense project (left) and a remix of that that project (right).

Answering RQ2, we find support for claims that remixing is more effective for code-heavy projects than for media-heavy works. Across all models, the interaction between log *blocks* and *isremix* is positive, although relatively small in magnitude. The interaction between log *media* and *isremix* is negative and strongly statistically significant in both models and is much larger than the interaction term with *blocks*.

The left panel of Figure 3 visualizes predicted loveits across varying code complexity while holding media complexity at its sample median of 10. The graph visualizes Model 1's prediction that *de novo* projects will receive more loveits than remixes at all levels of code complexity. That said, as projects become more complex, the predicted gap between remixes and *de novo* projects narrows slightly.

The difference between the effects of code-intensity and media-intensity for remixes is stark: all but the least media-intensive remix projects are predicted to receive more loveits than similar *de novo* projects. An otherwise average project with a single media element is predicted to receive 0.14 loveits – slightly higher than the estimate for an identical *de novo* project (0.11). However, if the same project contained 44 pieces of media (90[th] percentile), it would be expected to receive 0.18 loveits – much lower than the estimate of 0.35 loveits for an otherwise similar *de novo* work.

As predicted, we find that *views* is a strong predictor of *loveits*. Parameter estimates associated with log *views* are highly statistically significant in Model 2 and larger in magnitude than any other parameter. Controlling for *views* results in a substantial increase in goodness of fit and in large decreases in the magnitude of most of the other parameter estimates; as predicted, loveits is strongly influenced by exposure. That said, the essential pattern of results – in both the signs of parameter estimates of our key question predictors and their relative magnitude – is stable across the two models. In fact, controlling for views even strengthens aspects of our findings. For example, Model 2 predicts that the effects of *loveits* will always be lower for *media* than for *blocks* in remixes, and will decrease monotonically as remixes increase in media-intensity if other variables are held constant.

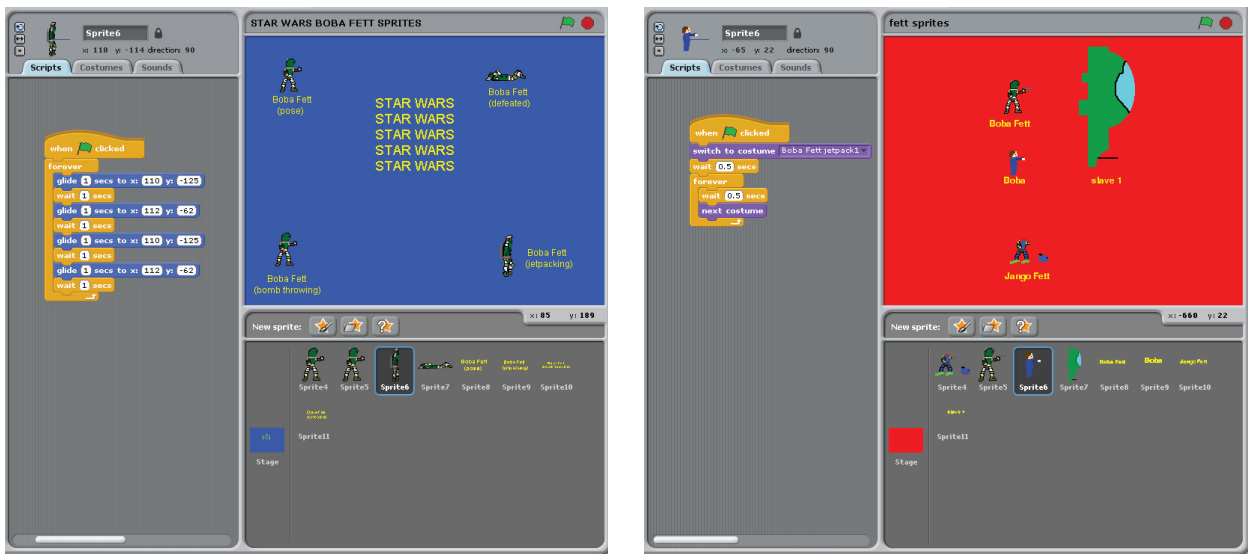

Figure 5: Example of a media-intense project (left) and a remix of that project (right).

Our controls are also statistically significant predictors of loveits. Surprisingly, we find that find that projects by older users and users with longer tenure on the site receive fewer loveits on average – although both of these effects are extremely small in magnitude. We find that our measure of prominence (*prevloveits*) is a powerful predictor of ratings but is largely attenuated by the addition of a control for exposure; more prominent users receive more loveits because they receive more attention. Finally, controlling for exposure changes a non-significant estimate of the parameter associated with female authorship into a large, positive, and statistically significant effect. We interpret this as suggesting that girls' projects are given more positive ratings, on average, but only after we take into account the fact that they tend to be viewed less often.

These findings can be made more concrete using two examples. The first example is shown in Figure 4. In the left panel of the figure is a "simple level editing system" created, *de novo*, by a 15-year-old American boy. The project has a framework to create a game with multiple platforms. In the right panel is a remix of that project created by a 13-year-old American boy who built a full-fledged game from the original project. The remixed project had 247 blocks (near the $90^{th}$ percentile of code intensity) but is near the sample median in media intensity. As our findings suggest happens on average, the code-intense remix received more loveits than its antecedent.

That example can be contrasted with a pair of media intense projects shown in Figure 5. On the left side is, once again, a *de novo* work: a project by 11-year-old boy that contained several animations of a Star Wars character that moves around the screen. A remixed version of the project, shown on the right, was created by an adult Scratch user to change the color and to replace and change a number of the the graphical elements. The remix project, very similar to its antecedent, contained 38 images (near the $90^{th}$ percentile in terms of media intensity) but near the sample median in terms of code. Again, as our findings suggest happens on average in Scratch, the remix of the the media-intense project was less highly rated, receiving no loveits all, while the *de novo* project received several.

In summary: Speaking to RQ1, we find surprising evidence that remixed works are less positively rated in Scratch, on average, than similar *de novo* projects for all but the least media-intensive works. Remixes are, in general, rated less highly than the works of single authors. In answer to RQ2, we find support for the common wisdom about the relatively poor quality of remixed art. Our results suggest that remixes are more positively rated when they are code-intensive than when they are media-intensive. We find that these basic associations are attenuated by, but robust to, consideration of the effect of exposure.

## LIMITATIONS

A number of important limitations and threats may affect the validity of our findings. One important concern that we have already alluded to is around our measure of quality. Most crowdsourcing and peer production research has focused on works with relatively clear and objective means of evaluation. Because we are interested in comparing works with artistic components, evaluation must be subjective. Other valid measures of quality – e.g., expert ratings – might result in different findings.

Our discussion of code and media intensity in Scratch in terms of functional versus artistic modes of production involves a series of important limitations and threats. Scratch users can use code to generate aesthetic or artistic work. Similarly, a very media intense project might serve a very functional purpose. Our independent variables measure the material of production – code and media – and do not capture conceptual models of creative production. The relationship between the tools and materials of production and the styles of products that result is a deep topic beyond the scope of this analysis. Our findings should be seen as limited in this regard.

Additionally, our empirical strategy assumes that media-intensive collaborative projects result from collaboration on the media itself. This may not always be the case. For example, a media-intensive work might be remixed entirely through addition of, or changes to, code. In that situation, the collaborative act would be collaboration on code but would be measured, in our dataset, merely as the amount of code and media in the resulting remix. Through our experience with Scratch and conversations with its administrators, we believe that remixed works will be similar to their antecedents and will involve collaboration distributed approximately proportionally across their different parts and types of content. However, we cannot reject this threat.

Previous work has shown that explicit and public expression of admiration may create self-

15reinforcing social dynamics [37]. A second measure, *favorites*, is a count of the number of times that a project has been listed by other users as a favorite – designed as a form of personal bookmarking. Like loveits, users can only favorite a given project once and a project being listed as a favorite is visible to both the project creator and the person "favoriting." However, unlike *loveits*, *favorites* are not displayed on project pages and are only visible by browsing to the bottom of the favoriter's profile page. As a result, this second measure is less likely to provoke concerns of self-reinforcing behaviors. Although the median number of favorites after one year is even lower than *loveits* (85% of projects are never marked as a favorite by another Scratch user), our basic pattern of results is similar when we use *favorites* as a dependent variable.

Another important threat is that our results may be driven by remixed projects competing for ratings with their antecedents. This may be the case because, although loveits are not limited, users' attention is. For example, if a game is lightly remixed and shared, there would then be two similar games in the community which would divide attention; each project would receive only a portion of the loveits that would otherwise have been given to the original. Our models controlling for exposure should help address this threat. We can gain additional confidence that this threat is not driving our results because remixed projects in our data receive many more loveits ($\mu = 4.6$) than projects which are never remixed ($\mu = 0.5$) and the difference is strongly statistically significant ($t = 46$; $p < 0.001$).

Another important limitation, common to any in-depth analysis of a single community, is generalizability beyond the empirical setting. Although our analysis only considers collaboration through remixing – one of several collaborative modalities – remixing is widely popular in contemporary youth culture. At the very least, multimedia remixing is widely practiced, in very similar forms, in other communities like Newgrounds and ccMixter [20, 18, 6, 29]. And although Scratch is a single site, it contains the works of over a million creators engaged in a wide variety of collaborative interactions. It is more difficult to speculate on how these results will generalize to adults, to other types of collaborative communities, or to different types of creative production. These remain open questions for future research.

Nevertheless, Scratch is unusual in several ways that inform questions of generalizability. For example, we can address our second research question because Scratch projects fluidly, and by design, mix artistic and functional elements in a single interface and community. This also means that Scratch projects exist in a genre that defies easy categorization. Scratch projects are rarely as purely artistic as a novel or painting and are rarely as purely functional as operating system software. Our findings should be seen as limited to this boundary area between functional and artistic works.

Although we argue that loveits are proxies for project quality, it is important to realize that loveits are constituted as social actions within a community of creators, evaluators, and collaborators. It is important to consider that different types of remixing in Scratch may also lead to variation in the norms around appropriate expression of appreciation between sub-communities.

By bringing together a wide variety of activities, Scratch provides a unique empirical setting that makes answering our research questions possible. However, the richness, breadth, and size of Scratch comes with a cost that makes our answers necessarily contingent. We hope that future work on these issues will explore and test our findings, and these limitations, in the growing number of similar peer production communities and beyond.

## DISCUSSION

Our paper contributes to the literature on peer production, remixing, and cooperative work in two key ways. First, we provide empirical evidence from a remixing community that collaborative projects are, on average, rated as being of lower quality than works of single authorship. This finding is surprising, and suggests new and important limits to very broad claims of the benefit of collaboration in remixing and peer production. Second, by using a unique dataset of content that mixes code and traditionally artistic media, we provide what we believe is the first empirical support for the common wisdom that collaboration may be more effective for functional works than for artistic media.

Our quality metric is based on community assessment but it is unclear when designers should care what a community thinks about a member's work. Remixing may be good *a priori*, if it encourages learning, promotes democratization of cultural production, or encourages problem solving. For example, the quality of remixes in Scratch might not be important if it allows people to learn and socialize. Similarly, even if remixing results in low quality products, it might, as Lessig and Benkler suggest, play a democratizing function for culture. Furthermore, the quality of remixes might be irrelevant to those innovators that are driven by a desire to solve their own problems, as theories of user innovation suggest [41]. While high quality products have been an important driver of interest in peer production, peer evaluations of quality may be less important or relevant to system designers when considered from all three of these perspectives.

Both Lanier and Keen criticize remixing and peer production not only for producing low quality work, but also for lacking originality and the ability to produce real innovation. Even Raymond raises questions about the possibility that peer production can create innovative products and suggests that, fundamentally, "insight comes from individuals" [35]. Although previous work on Scratch has shown evidence of a wide range of derivative and transformative remixes [16], the innovativeness of a given project is both difficult to ascertain and fundamentally subjective. And although we know that less original remixes result in less happy users [17], it is difficult to know to what degree user ratings reflect novelty, creativity, or the spark of innovation. Finally, although it seems reasonable to suggest that the average remix will be less innovative – by virtue of being a remix – than a similar *de novo* project, interest in remixing stems, in part, from the fundamental Schumpeterian concept of innovation through recombination [38].

Our results suggest that people rate code-based remixes more positively that media-based remixes. One possible explanation is that Scratch is more modular for code than for media. As a result, remixing of media may lead to a struggle to maintain a consistency of vision or execution in way that code-based remixing does not. This decreased consistency may be reflected in lower ratings. In broader terms, perhaps the Scratch platform is simply not well designed for artistic collaboration. As a first step, designers can address this by building systems that support more modular media to support more structured collaboration.

A second explanation is that people perceive collaboration in code in terms of objective improvements while media elements will tend to be viewed more subjectively. This subjectivity in the evaluation of projects could lead to less positive assessment on average. Although it represents a challenge, future work could address both explanations by comparing different systems and considering the structure and shape of this variation.

The creation of richer avenues for productive interaction between media and code is an important area for further work in Scratch and elsewhere. Our results suggest that collaborative creativity, already cited in the HCI literature as an important and challenging area for future research, is likely to face particularly stark challenges for more media-intensive works.

What we hope to offer in this paper is a step toward a fuller understanding of the limitations of collaboration in remixing and peer production. As others have suggested and shown, collaboration is not *always* better. In Scratch, at least, it usually results in lower peer ratings of the resulting products – particularly for media-intensive works. As supporters and advocates of remixing, we reject the blind repetition of the mantra that collaboration is better for all types of work. Instead, we believe it is crucial to learn and understand the limitations and challenges associated with these important cultural practices. With this knowledge, we hope to more effectively influence the design of social media and collaboration support systems, and to help peer production thrive.

## ACKNOWLEDGMENTS

We thank the Lifelong Kindergarten Group at the MIT Media Laboratory for creating Scratch as well as the thousands of children and adult members of the Scratch online community. Scratch is a project with financial support from the National Science Foundation (Grant No. ITR-0325828), Microsoft Corp., Intel Foundation, Nokia, and the MIT Media Lab research consortia. Work in this paper was funded by the MIT Sloan School of Management, the MIT Media Lab Consortia, Microsoft Research, the Berkman Center for Internet and Society at Harvard University, NSF grants OCI-1026473 and OCI-1027848 and the Ford Foundation's Visionary Leadership gift.